\documentclass[prl,aps,twocolumn,showpacs,preprintnumbers,amsmath,amssymb]{revtex4}
\usepackage[dvips]{graphicx}
\usepackage[]{caption}
\usepackage{amsmath}
\usepackage{amssymb}

\usepackage{graphicx}
\usepackage{dcolumn}
\usepackage{bm}
\def\be{\begin{equation}}
\def\ee{\end{equation}}
\def\bea{\begin{eqnarray}}
\def\eea{\end{eqnarray}}
\newcommand{\pp}{\Psi^{\prime}} 
 
\newcommand{\cp}{\chi^{\prime}} 
\newcommand{\fp}{\Phi^{\prime}} 
\newcommand{\vp}{\phi^{\prime}}

\newcommand{\ppp}{\Psi^{\prime\prime}} 
 
\newcommand{\cpp}{\chi^{\prime\prime}} 
 
\newcommand{\vpp}{\phi^{\prime\prime}} 
 
\newcommand{\ch}{{\cal{H}}} 
 
\begin{document}

\preprint{DCTP-06/21,HUTP-06/A0035}

\date{\today}

\title{More on the Spectrum of Perturbations in String Gas Cosmology}

\author{Robert H. Brandenberger$^{1)}$} \email[email: ]{rhb@hep.physics.mcgill.ca}
\author{Sugumi Kanno$^{1)}$} \email[email: ]{sugumi@hep.physics.mcgill.ca}
\author{Jiro Soda $^{2)}$} 
\email[email: ]{jiro@tap.scphys.kyoto-u.ac.jp}
\author{Damien A. Easson$^{3)}$} \email[email: ]{damien.easson@durham.ac.uk}
\author{Justin Khoury$^{4)}$} \email[email: ]
{jkhoury@perimeterinstitute.ca}
\author{Patrick Martineau$^{1)}$} \email[email: ]
{martineau@hep.physics.mcgill.ca}
\author{Ali Nayeri$^{5)}$} \email[email: ]{nayeri@feynman.harvard.edu}
\author{Subodh P. Patil$^{1)}$} \email[email: ]{patil@hep.physics.mcgill.ca}

\affiliation{1) Department of Physics, McGill University, Montr\'eal, QC, H3A 2T8, Canada}
\affiliation{2) Department of Physics, Kyoto University, Kyoto, Japan}
\affiliation{3) Centre for Particle Theory, Department of
Mathematical Sciences, Durham University, Science Laboratories, South Road, Durham, DH1 3LE, U.K.}
\affiliation{4) Perimeter Institute, 31 Caroline St. N., Waterloo, ON,
N2L 2Y5, Canada}
\affiliation{5) Jefferson Physical Laboratory, Harvard University,
Cambridge, MA, 02138, USA}


\begin{abstract}

String gas cosmology is rewritten in the Einstein frame. In
an effective theory in which a gas of 
closed strings is coupled to a dilaton
gravity background without any potential for the dilaton,
the Hagedorn phase which is quasi-static in the string frame corresponds
to an expanding, non-accelerating phase from the point of view of
the Einstein frame. The Einstein frame curvature singularity which appears 
in this toy model is related to the blowing up
of the dilaton in the string frame. However, for large values of
the dilaton, the toy model clearly is inapplicable. Thus, there
must be a new string phase which is likely
to be static with frozen dilaton. With such a phase, the horizon
problem can be successfully addressed in string gas cosmology. The
generation of cosmological perturbations in the Hagedorn phase
seeded by a gas of long strings in thermal equilibrium is reconsidered,
both from the point of view of the string frame (in which it is easier
to understand the generation of fluctuations) and the Einstein 
frame (in which the evolution equations are well known). It is
shown that fixing the dilaton at some early stage is important
in order to obtain a scale-invariant
spectrum of cosmological fluctuations in string gas cosmology. 

\end{abstract}

\maketitle

\newcommand{\eq}[2]{\begin{equation}\label{#1}{#2}\end{equation}}

\section{Introduction}
  
String gas cosmology is a model of superstring cosmology which is
based on coupling to a classical dilaton gravity background
a gas of classical strings with a mass spectrum
corresponding to one of the consistent perturbative superstring
theories \cite{BV,TV} (see also \cite{Perlt} for
early work, and \cite{RHBrev1,RHBrev2,BattWat} for reviews). String gas
cosmology has been developed in some detail in recent years
\cite{ABE,BEK,others}. In particular, it was shown 
\cite{Watson,Patil1,Patil2,Edna,Watson2,Sugumi,other2} that
string modes which become massless at enhanced symmetry
points lead to a stabilization of the volume and shape moduli of
the six extra spatial dimensions (see \cite{BV,Mairi} for
arguments in the context of string gas cosmology on how
to naturally obtain the separation between three large and
six string-scale dimensions \footnote{See, however, 
\cite{Col,Danos} for some caveats.}) .

String gas cosmology is usually formulated in the string frame,
the frame in which stringy matter couples canonically to the
background dilaton space-time. The existence of a maximal
temperature \cite{Hagedorn} of a gas of 
weakly interacting strings in thermal
equilibrium has crucial consequences for string cosmology. 
As discussed in \cite{BV}, as we follow
our universe back in time through the radiation phase of standard
cosmology, then when the temperature approaches its limiting value,
the energy shifts from the radiative modes to the string oscillatory
and winding modes. Thus, the pressure approaches zero. In the
string frame, and for zero pressure,
as follows from the equations derived in \cite{TV,Ven},
the universe is quasi-static. There is an attractive fixed point
of the dynamics in which the scale factor of our large three
dimensions is constant, but the dilaton is dynamical (we consider
the branch of solutions in which the dilaton is a decreasing function
of time). We call this phase the quasi-static Hagedorn phase. 
Since the string frame Hubble radius is extremely large in the
Hagedorn phase (infinite in the limiting case that the scale factor
is exactly constant), but decreases dramatically during the
transition to the radiation phase of standard cosmology, all
comoving scales of interest in cosmology today are sub-Hubble
initially, propagate on super-Hubble scales for a long time after
the transition to the radiation phase before re-entering the
Hubble radius at late times. Thus, it appears in principle possible
to imagine a structure formation mechanism driven by local physics.

As was recently suggested \cite{ABV}, string thermodynamic fluctuations
in the Hagedorn phase may lead to a nearly scale-invariant spectrum of
cosmological perturbations. There would be a slight red tilt for the
spectrum of scalar metric perturbations. A key signature of this
scenario would be a slight blue tilt for the spectrum of gravitational
waves \cite{BNPV} (see also \cite{BNPV2,Ali} for more detailed
treatments). Since this result is surprising from the point of view
of particle cosmology, and since the analyses of \cite{ABV,BNPV}
contained approximations, it is an interesting challenge to analyze 
the cosmology from the point of view of the Einstein frame, the frame in
which cosmologists have a better physical intuition.

In this paper, we analyze both the background dynamics of string gas
cosmology and the generation and evolution of cosmological perturbations
in the Einstein frame. A naive extrapolation of the background solutions
of \cite{TV} into the past would yield a cosmological singularity.
Such an extrapolation is clearly not justified once the dilaton
reaches values for which we enter the strong coupling regime of
string theory. Rather, at early times one must have a new phase
of the theory in which the dynamics    is consistent with the
qualitative picture which emerges from string thermodynamics.   
This phase is meta-stable and will decay into a phase
with rolling dilaton. The modified background evolution
can solve the horizon and singularity problems in the context 
of string gas cosmology. 

An improved analysis of generation of fluctuations in the
string frame shows that the conclusions of
\cite{ABV,BNPV} concerning the spectra of cosmological perturbations and
gravitational waves 
(we are considering the case where our three large dimensions
are toroidal) are only obtained if the dilaton velocity can be 
neglected.

\section{Background Dynamics in the Einstein Frame}

String gas cosmology is based on T-duality symmetry and on
string thermodynamics. String thermodynamics yields the existence
of a maximal temperature of a gas of strings in thermal
equilibrium, the Hagedorn temperature \cite{Hagedorn}. If we
consider \cite{BV} adiabatic evolution of a gas of strings in
thermal equilibrium as a function of the radius of space $R$,
T-duality \cite{tduality} yields a temperature-radius curve
(see Fig. 1) which is symmetric about the self-dual radius. Being at
the self-dual radius must be a fixed point of the dynamics. The higher
the entropy of the string gas is at a fixed radius, the
larger is the flat region of the curve, the region where
the temperature remains close to the Hagedorn temperature. 
This implies that the duration of the Hagedorn phase will increase
the larger the energy density is. 

\begin{figure}
\includegraphics[height=6cm]{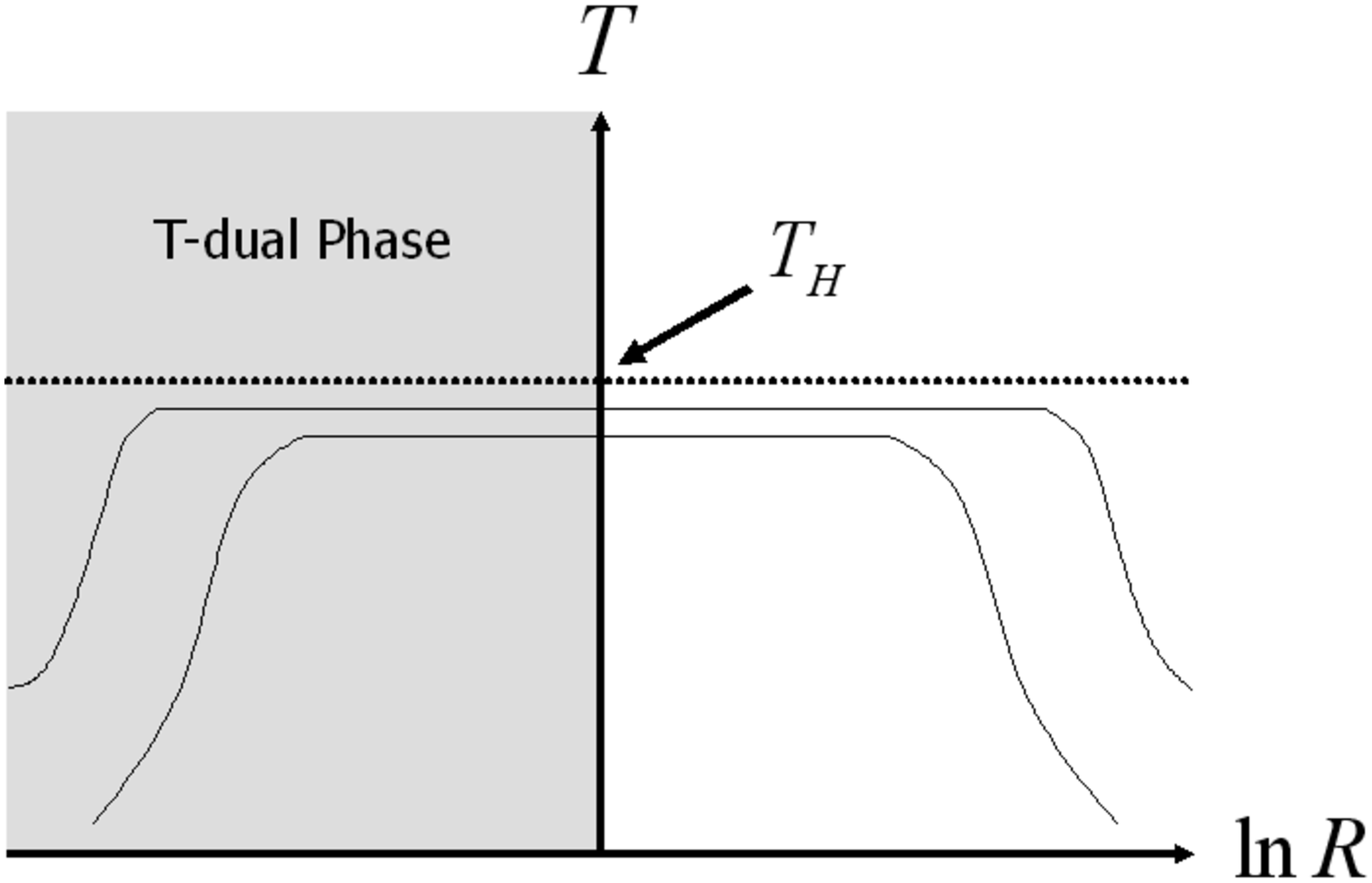}
\begin{caption}
{\small Sketch (based on the analysis of \cite{BV}
of the evolution of temperature $T$ as a function
of the radius $R$ of space of a gas of strings in thermal
equilibrium. The top curve is characterized by an entropy
higher than the bottom curve, and leads to a longer
region of Hagedorn behaviour.}
\end{caption}
\label{fig:1}
\end{figure}

In a regime in which it is justified to consider the dynamics in
terms of a gas of strings coupled to background dilaton gravity, the 
string frame action for the background fields, the metric and
the dilaton (we will set the antisymmetric tensor field to zero) is
\be \label{E0}
S \, = \, - \int d^{1 + N}x \sqrt{-g}e^{-2\phi}
\bigl[R + 4 g^{\mu \nu} \partial_{\mu} \phi \partial_{\nu} \phi \bigr] \, ,
\ee
where $R$ is the string frame Ricci scalar, $g$ is the determinant of
the string frame metric, $N$ is the number of spatial dimensions,
and $\phi$ is the dilaton field. Note that we are working in units
in which the dimensionful  pre-factor appearing in front of the
action is set to $1$. 

The background is sourced by
a thermal gas of strings. Its action $S_m$ is given by the
string gas free energy density $f$ (which depends on the string
frame metric) via
\be
S_m \, = \, \int d^{1 + N}x \sqrt{-g} f \, .
\ee
The total action is the sum of $S$ and $S_m$. Note that the factor
$e^{-2\phi}$ gives the value of Newton's gravitational constant.

In the case of a spatially flat, homogeneous
and isotropic background given by
\be \label{metric}
ds^2 \, = \, dt^2 - a(t)^2 d{\bf x}^2 \, ,
\ee
the three resulting equations of motion of dilaton-gravity (the
generalization of the two Friedmann equations plus the equation
for the dilaton) in the string frame are
\cite{TV} (see also \cite{Ven})
\bea
-N {\dot \lambda}^2 + {\dot \varphi}^2 \, &=& \, e^{\varphi} E 
\label{E1} \\
{\ddot \lambda} - {\dot \varphi} {\dot \lambda} \, &=& \,
{1 \over 2} e^{\varphi} P \label{E2} \\
{\ddot \varphi} - N {\dot \lambda}^2 \, &=& \, {1 \over 2} e^{\varphi} E \, ,
\label{E3}
\eea
where $E$ and $P$ denote the total energy and pressure, respectively,
and we have introduced the logarithm of the scale factor 
\be
\lambda(t) \, = \, {\rm log} (a(t))
\ee
and the rescaled dilaton
\be
\varphi \, = \, 2 \phi - N \lambda \, .
\ee
where $N$ is the number of spatial dimensions (in the following, 
the case of $N = 3$ will be considered).

The Hagedorn phase is characterized by vanishing $P$ and, therefore, 
constant total energy $E$. Thus, 
combining (\ref{E1}) and (\ref{E3}) to eliminate the dependence on
${\dot{\lambda}}$, yields a 
second order differential equation for $\varphi$ with the solution
\be \label{E4}
e^{-\varphi(t)} \, = \, {{E_0} \over 4} t^2 - 
{\dot{\varphi_0}} e^{-{\varphi_0}} t + e^{-{\varphi_0}}
\ee
subject to the initial condition constraint
\be \label{E5}
{\dot{\varphi_0}}^2 \, = \, e^{\varphi_0} E_0 + N {\dot{\lambda_0}}^2
\ee
which follows immediately from (\ref{E1}). In the above, the subscripts
stand for the initial values at the time $t = 0$. 

In the Hagedorn phase, the second order differential equation 
(\ref{E2}) for
$\lambda$ can easily be solved. If the initial conditions require
non-vanishing $\dot{\lambda_0}$, the solution is
\be \label{E6}
\lambda(t) \, = \, \lambda_0 + {1 \over {\sqrt{N}}}
{\rm ln} \bigl[ {{{\sqrt{N}} {\dot{\lambda_0}} - {\cal G}} \over 
{{\sqrt{N}} {\dot{\lambda_0}} + {\cal G}}}  \bigr] \, ,
\ee
where ${\cal G}$ is an abbreviation which stands for
\be
{\cal G} \, = \, {{E t} \over 2} e^{\varphi_0} - {\dot{\varphi_0}} \, .
\ee

{F}or vanishing initial value of the derivative of the scale factor, the
solution is simply
\be \label{E7}
\lambda(t) \, = \, \lambda_0 \, ,
\ee
i.e. a static metric. In the static case, the result (\ref{E4})
simplifies to
\be \label{E8}
e^{-\varphi(t)} \, = \, e^{-\varphi_0} 
\bigl({{{\dot{\varphi_0}} t} \over 2} - 1 \bigr)^2 \, .
\ee
These solutions are slight generalizations of the solutions given
in the appendix of \cite{TV}. These solutions have also very
recently been discussed in \cite{TK}. We are interested in
the branch of solutions with $\dot{\phi} < 0$.

One important lesson which follows from the above solution is that,
although the metric is static in the Hagedorn phase, a dilaton 
singularity develops at a time $t_s$ given by
\be \label{E9}
t_s \, = \, - {2 \over {|{\dot{\varphi_0}}|}} \, .
\ee
In fact, already at a slightly larger time $t_{c}$, 
the dilaton has reached the critical value $\phi = 0$, beyond which 
string perturbation theory breaks down. The times $|t_s|$ and
$|t_{c}|$ are typically of string scale. Thus, unless the
current value of the dilaton is extremely small, the duration of the
phase in which the above solution is applicable will be short.

Note, however, that the solution (\ref{E8}) is not consistent
with the qualitative picture which emerges from string thermodynamics
\cite{BV} (see Fig. 1) according to which the evolution of all
fields close to the Hagedorn temperature should be almost static.
We know that the dilaton gravity action ceases to be justified in
the region in which the theory is strongly coupled. This leads to the
conclusion that the phase during which (\ref{E8}) is applicable must
be preceded by another phase of Hagedorn density, a phase in which
the dynamics reflects the qualitative picture which emerges from
Fig. 1, and corresponds to fixed
scale factor and fixed dilaton. We call this phase the {\it strong coupling
Hagedorn phase} \footnote{Another argument supporting the
assumption that in the strong coupling Hagedorn phase the dilaton
is fixed can be given making use of S-duality. Under S-duality,
the dilaton $\phi$ is mapped to $- \phi$. It is reasonable to
assume that close to the maximal temperature, the system is in a 
configuration which is self S-dual, and in which the dilaton is hence
fixed - we thank C. Vafa for stressing this point. Note that we
are assuming that the existence of the maximal temperature remains
true at strong coupling.}
Note that the Einstein action is not invariant under
T-duality. Hence, we expect that intuition based on Einstein gravity
will give very misleading conclusions when applied to the
strong coupling Hagedorn phase. In particular, constant energy 
density should {\it not}
lead to a tendency to expansion. The strong coupling phase
is long-lived but meta-stable and decays into a solution
in which the dilaton is free to roll, a phase described by
the equations (\ref{E1} - \ref{E3}).

If the equation of state is that of radiation, namely 
$P = 1/N E$, then a solution with static dilaton is an attractor.
For static dilaton, the equations (\ref{E2}) and (\ref{E3})
then reduce to the usual Friedmann-Robertson-Walker-Lemaitre
equations.

\begin{figure}
\includegraphics[height=6cm]{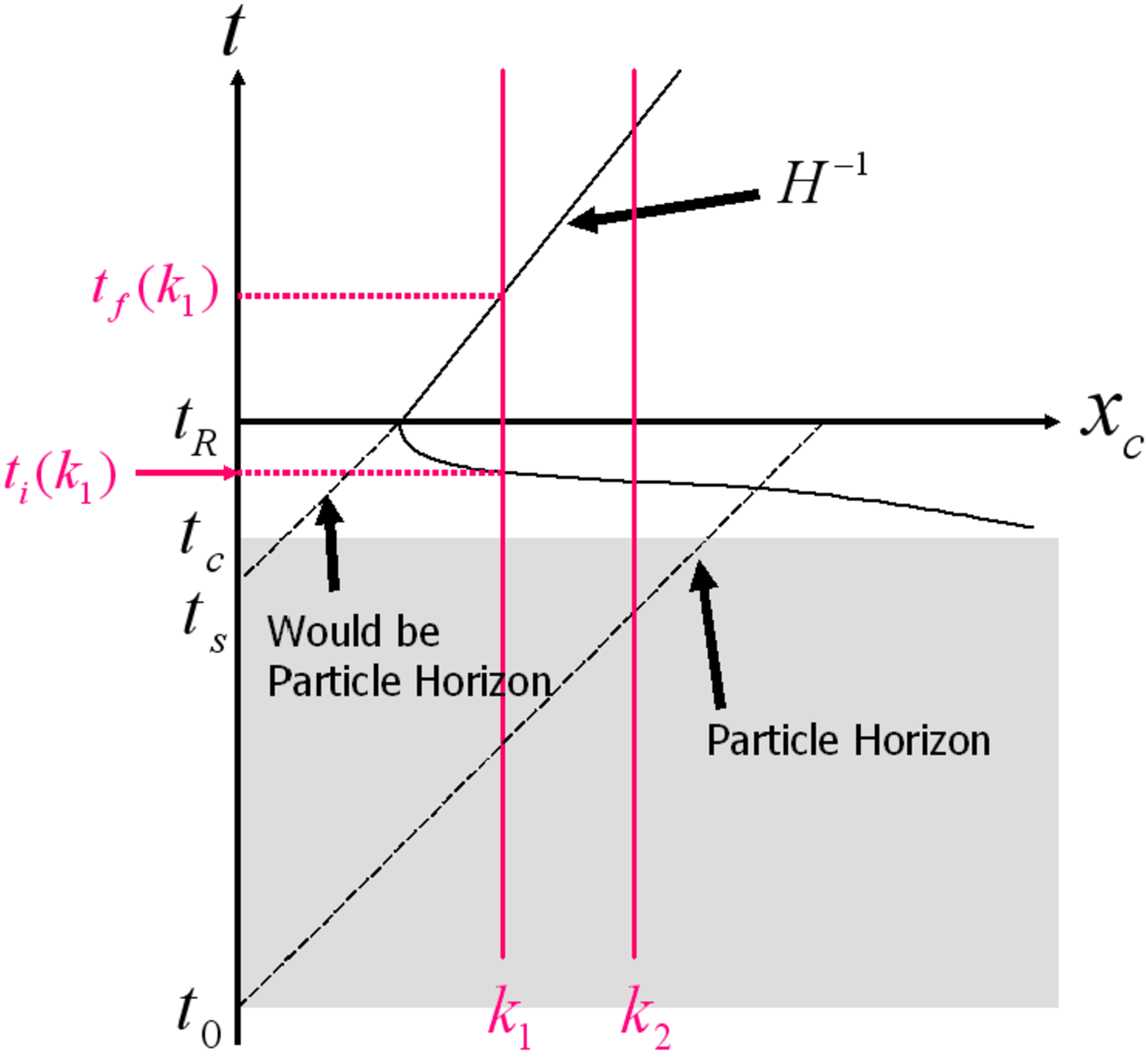}
\begin{caption}
{\small Space-time diagram (sketch) showing the evolution
of fixed comoving scales in string gas cosmology. 
The vertical axis is string frame time, the horizontal
axis is comoving distance. The Hagedorn phase ends at the time $t_R$
and is followed by the radiation-dominated phase of standard
cosmology. The solid curve represents the Hubble radius $H^{-1}$
which is cosmological during the quasi-static Hagedorn phase,
shrinks abruptly to a micro-physical scale at $t_R$ and then
increases linearly in time for $t > t_R$. Fixed comoving scales
(the dotted lines labeled by $k_1$ and $k_2$) which are currently probed in
cosmological observations have wavelengths which are smaller than
the Hubble radius during the Hagedorn phase. They exit the 
Hubble radius at
times $t_i(k)$ just prior to $t_R$, and propagate with a wavelength
larger than the Hubble radius until they reenter the Hubble radius
at times $t_f(k)$.  
  Blindly extrapolating the solutions (\ref{E7}, \ref{E8})
into the past would yield a dilaton singularity
at a finite string time distance in the past of $t_R$, at a time
denoted $t_s$. However, before this time is reached (namely
at time $t_c$) a transition
to a strong coupling Hagedorn phase with static dilaton is reached. Taking
the initial time in the Hagedorn phase to be $t_0$, the forward
light cone from that time on is shown as a dashed line. 
The shaded region corresponds to the strong coupling Hagedorn phase. 
}
\end{caption}
\label{fig:2}
\end{figure}

{F}igure 2 shows a space-time sketch from the perspective of
string frame coordinates. The Hagedorn phase lasts until 
the time $t_R$ (the time interval from $t_c$ to close to $t_R$ being
describable by the equations of motion of dilaton gravity)
when a smooth transition to the
radiation phase of standard cosmology takes place. This transition
is governed by the annihilation of string winding modes into
oscillatory modes and is described by Boltzmann-type
equations \cite{BEK} (with corrections pointed out in \cite{Col,Danos}).
These Boltzmann equations are analogs of the equations used in
the cosmic string literature (see \cite{CS} for
reviews) to describe the transfer of energy between ``long" (i.e.
super-Hubble) strings and string loops. Note that the decay
of string winding modes into radiation is the process that ``reheats"
the universe.
  
Close to the time $t_s$, we reach the strong coupling Hagedorn phase,
the phase responsible for generating a large horizon.
  
In order that the cosmological background of Figure 2 match with our
present cosmological background, the radius of space at the end of
the Hagedorn phase needs to be of the order of $1$mm, 
the size that expands into our currently observed universe making use
of standard cosmology evolution beginning at a temperature of
about $10^{15}$GeV. This is many
orders of magnitude larger than the string size. Thus, without
further assumptions, there is a cosmological ``horizon'' and 
``entropy'' problem, similar
to the one present in Standard Big Bang (SBB) cosmology. Provided
that the strong coupling Hagedorn phase is long-lived (and, based on 
Fig. 2 this
is more likely the higher the initial energy density is chosen), 
string gas cosmology will be able to resolve these problems. In
particular, there will be enough time to establish  thermal equilibrium 
over the entire spatial section, a necessary condition for the
structure formation scenario outlined in \cite{ABV} to work.

The above issues become more manifest when the cosmological
background is rewritten in the Einstein frame. It is to this subject to which
we now turn.

The conformal transformation of the metric to the Einstein frame is 
given by
\be \label{E10}
{\tilde{g}}_{\mu \nu} \, = \, e^{- 4 \phi / (N - 1)} g_{\mu \nu} \,
= \, e^{-2 \phi} g_{\mu \nu}
\ee
where quantities with a tilde refer to those in the Einstein frame,
and in the final expression we have set $N = 3$.
The dilaton transforms as (for $N = 3$)
\be \label{E11}
{\tilde \phi} \, = \,  2 \phi \, .
\ee
Under this transformation, the action (\ref{E0}) becomes
\be \label{E12}
S_E \, = \, - \int d^4x \sqrt{-{\tilde{g}}}
\bigl( {\tilde{R}} - {1 \over 2} {\tilde{g}}^{\mu \nu}
{\tilde{\nabla}_{\mu}}{\tilde{\phi}} {\tilde{\nabla}_{\nu}}{\tilde{\phi}}
\bigr) \, .
\ee
The matter action in the Einstein frame becomes
\be
S_m \, = \,  \int d^4x e^{2 \phi} f[\tilde g, \phi] \sqrt{\tilde g} \, ,
\ee
from which we see that the factor $e^{2 \phi}$ plays the role of
the gravitational constant.

We apply the conformal transformation for the metric (\ref{metric})
of a homogeneous and isotropic universe. To put the resulting
Einstein frame metric into the FRWL form, we have to re-scale the
time coordinate, defining a new Einstein frame time ${\tilde{t}}$
via
\be \label{E13}
d{\tilde{t}} \, = \, e^{- \phi} dt \, .
\ee
The resulting scale factor ${\tilde{a}}$ in the Einstein frame
then is given by
\be \label{E14}
{\tilde{a}} \, = \, e^{- \phi} a \, .
\ee
The comoving spatial coordinates are unchanged.

The Hubble parameters ${\tilde{H}}$ and $H$ in the Einstein and
string frames, respectively, are related via
\be \label{E15}
{\tilde{H}} \, = \, e^{\phi} \bigl(H - {\dot{\phi}} \bigr) \, .
\ee
It is important to note that they denote very different lengths.

Let us now calculate the evolution of the Einstein frame
scale factor. Making use of the solution for the string frame
dilaton (\ref{E8}), it follows from (\ref{E14}) that
\be \label{sol1}
{\tilde{a}} \, = \, e^{\lambda_0} e^{- \phi_0} 
\bigl(1 - {\dot{\phi_0}} t \bigr) \, .
\ee
By integrating (\ref{E13}) we find that the physical time in the
Einstein frame is given by
\be \label{sol2}
{\tilde{t}} \, = \, e^{- \phi_0} t 
\bigl( 1 - {1 \over 2}{\dot{\phi_0}} t \bigr) \, .
\ee
It is important to keep in mind that ${\dot{\phi}} < 0$.

It follows that, in the Einstein frame, 
the evolution looks like that of a universe dominated by radiation. This
is easy to verify for large times, 
when the factors of $-1$ within the parentheses in (\ref{sol1})
and (\ref{sol2}) are negligible, and it thus it follows that
\be \label{sol3}
{\tilde{a}}({\tilde{t}}) \, \sim \, {\tilde{t}}^{1/2} \, .
\ee
The same conclusion can also be reached for times close to the
singularity, by explicitly inverting (\ref{sol2}) and
inserting into (\ref{sol1}).

We thus see that the expansion is non-accelerated and the
Hubble radius is expanding linearly. The dilaton singularity in
the string frame is translated into a curvature singularity in the
Einstein frame, a singularity which occurs at the Einstein frame
time
\be \label{sol5}
{\tilde{t}}_s \, = \, - {{exp(-\phi_0)} \over 2}{|\dot{\phi_0}|}^{-1} \, .
\ee
From the constraint equation (\ref{E5}) it follows that this    
of the order of the string scale.    

As stressed earlier, these solutions cannot be applied when
the value of the dilaton is larger than $0$ (when the string
theory enters the strong coupling regime). Instead, we will
have a strong coupling Hagedorn phase characterized by constant dilaton,
and hence also almost constant Einstein frame scale factor.

The space-time sketch of our cosmology in the Einstein frame is
sketched in Figure 3. In the absence of the
strong coupling Hagedorn phase, the horizon (forward light cone
beginning at ${\tilde{t}}_s$) would follow the Einstein frame Hubble
radius (up to an irrelevant factor of order unity),
thus yielding a horizon problem. However, during the
strong coupling Hagedorn phase and, in particular, during the 
transition between
the strong coupling Hagedorn phase and the phase described by the 
rolling dilaton,
the horizon expands to lengths far greater than the Einstein
frame Hubble radius.
This phase can, in particular, establish thermal equilibrium
on scales which are super-Hubble in the rolling dilaton phase.
In the last section of this paper we will come back to a
discussion of how to model the strong coupling Hagedorn phase.

\begin{figure}
\includegraphics[height=6cm]{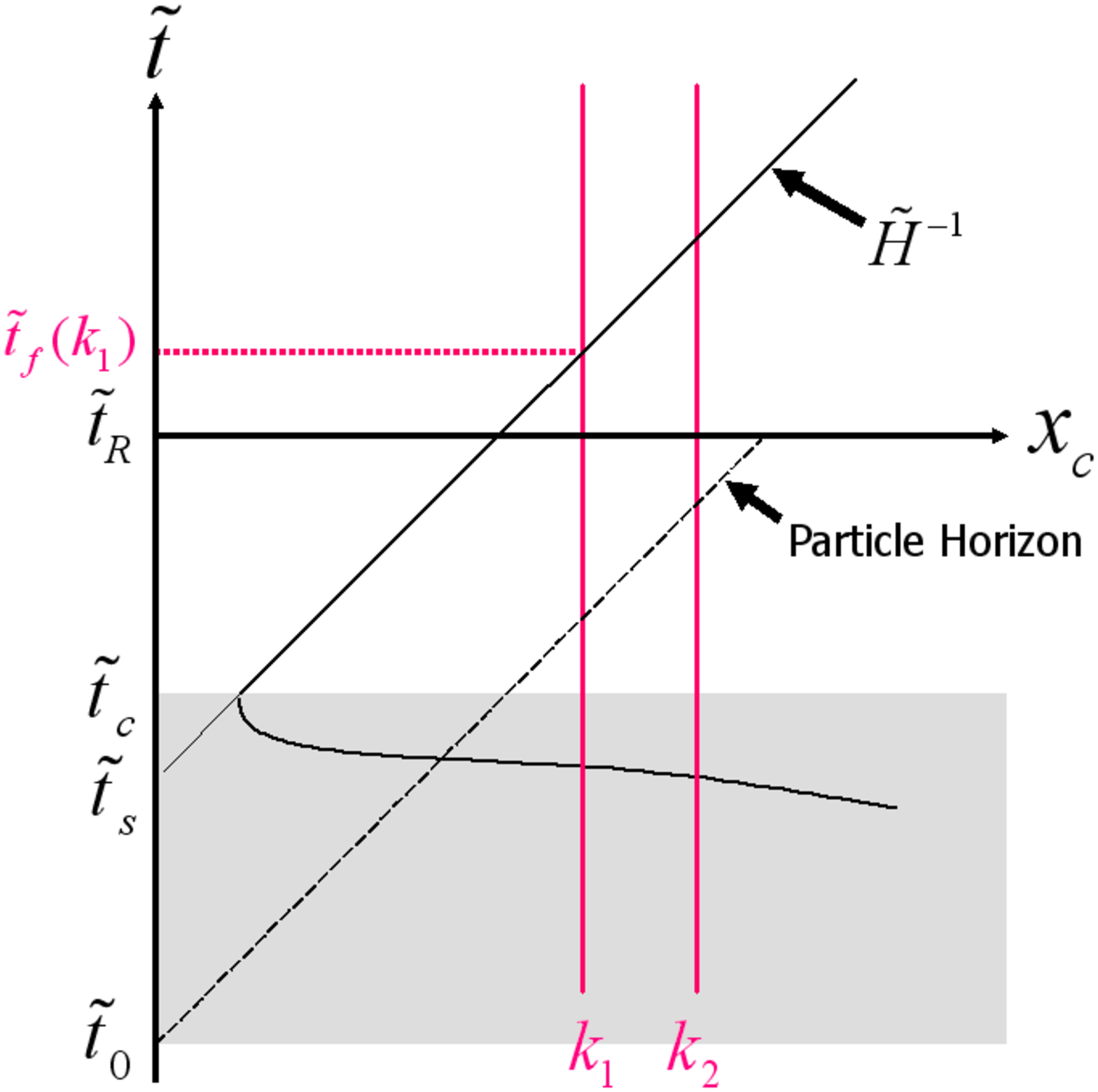}
\begin{caption}
{\small Space-time diagram (sketch) showing the evolution
of fixed comoving scales in string gas cosmology. 
The vertical axis is Einstein frame time, the horizontal
axis is comoving distance.  The solid curve represents the 
Einstein frame Hubble radius ${\tilde{H}}^{-1}$ which is linearly
increasing after ${\tilde{t}}_c$. Fixed comoving scales
(the dotted lines labeled by $k_1$ and $k_2$) which are currently probed in
cosmological observations have wavelengths which are larger than
the Einstein frame Hubble radius during the part of the Hagedorn phase
in which the dilaton is rolling. However, due to the presence of
the initial strong coupling Hagedorn phase, the horizon becomes much larger 
than
the Hubble radius. The shaded region corresponds to the strong coupling 
Hagedorn phase.}
\end{caption}
\label{fig:3}
\end{figure}

We will close this section with some general comments about the
relationship between the string and the Einstein frames. Obviously,
since the causal structure is unchanged by the conformal transformation,
the comoving horizon is frame-independent. The Hubble radius, on the
other hand, is a concept which depends on the frame. 
The string frame and the Einstein frame Hubble radii are two very 
different length scales. Both can be calculated
in any frame, but they have different meanings.
 
\section{Cosmological Perturbations in the String Frame}

From the point of view of the string frame, the scale of
cosmological fluctuations is sub-Hubble during the Hagedorn
phase. In \cite{ABV} it was assumed that thermal equilibrium
in the Hagedorn phase exists on scales of the order of $1$mm.
It was proposed to follow the string thermodynamical
matter fluctuations on a scale $k$ until that scale exits the
string frame Hubble radius at the end of the Hagedorn phase, to
determine the induced metric fluctuations at that time, and to
follow the latter until the present.

To study cosmological perturbations, we make use of a particular
gauge choice, longitudinal gauge (see \cite{MFB} for an in depth
review article on the theory of cosmological perturbations and
\cite{RHBrev4} for a pedagogical introduction), in which the metric
takes the form
\be \label{pertmetric}
ds^{2} \, = \, 
e^{2 \lambda(\eta)} \Bigl( (1+2\Phi) d\eta^{2}- (1-2\Psi)\delta_{i j}
dx^{i} dx^{j}\Bigr) \, ,
\ee
where $\eta$ is conformal time, and where $\Psi$ and $\Phi$ are the
fluctuation variables which depend on space and time. In Einstein
gravity, and for matter without anisotropic stress, the two potentials
$\Phi$ and $\Psi$ coincide, and $\Phi$ is the relativistic generalization
of the Newtonian gravitational potential. In the case of dilaton gravity,
the two potentials are related via the fluctuation of the dilaton field.

According to the proposal of \cite{ABV}, the cosmological perturbations are
sourced in the Hagedorn phase by string thermodynamical fluctuations, in
analogy to how in inflationary cosmology the quantum matter fluctuations
source metric inhomogeneities \footnote{In spite of this analogy,
there is a key difference, a difference which is in fact more important:
In inflationary cosmology, the fluctuations are quantum vacuum
perturbations, whereas in our scenario they are classical thermal
fluctuations.}. On sub-Hubble scales, the matter
fluctuations dominate. Hence it was suggested in \cite{ABV,BNPV2} to track
the matter fluctuations on a fixed comoving scale $k$ until the wavelength
exits the Hubble radius at time $t_i(k)$ (see Fig. 2). At that time,
the induced metric fluctuations are calculated making use of the
Poisson equation    
\be \label{Poisson}
\nabla^2 \Psi \, = \, 4 \pi G a^2 \delta T^0_0 \, .
\ee

The key feature of string thermodynamics used in \cite{ABV} (and discussed
in more detail in \cite{Ali,BNPV2}) 
is the fact that the specific heat $C_V$
scales as $R^2$, where $R$ is the size of the region in which we are
calculating the fluctuations. This result was derived in \cite{Deo}
and holds in the case of three large dimensions with topology of a
torus. The specific heat, in turn, determines
the fluctuations in the energy density. The scaling 
\be
C_V \, \sim \, R^2
\ee
leads to a Poisson spectrum
\be
P_{\delta \rho}(k) \, \sim \, k^4
\ee
for the energy density fluctuations, and a similar spectrum for the
pressure fluctuations \cite{BNPV,BNPV2}. Making use of the Poisson
equation (\ref{Poisson}), this leads to a scale-invariant spectrum
for the gravitational potential $\Phi$ \footnote{Note that
thermal particle fluctuations would not give rise to a
scale-invariant spectrum - see \cite{Joao} for an interesting
study of the role of thermal particle fluctuations in 
cosmological structure formation.}.

However, in the context of a relativistic theory of gravity, what
should be used to relate the matter and metric fluctuations
is the time-time component of the perturbed Einstein equation,
or - more specifically - its generalization to dilaton gravity,
and not simply the Poisson equation.
There are correction terms compared to (\ref{Poisson}) coming
from the expansion of the cosmological background and from the
dynamics of the dilaton. The analysis of \cite{ABV,BNPV} was done in 
the string frame. In this frame, during the Hagedorn phase the
correction terms coming from the expansion of space are not present,
but the dilaton velocity is important. The important concern is 
whether the resulting correction terms will change the conclusions
of the previous work. 

The perturbation equations in the string frame were discussed in
\cite{WB}. Denoting the fluctuation of the dilaton field by
$\chi$, the equations read
\bea \label{eq111}
{\vec \nabla}^2 \Psi   &-& 3\ch \pp - 3 \ch^2 \Phi \\ 
&=& \frac{1}{2} e^{2\phi + 2\lambda}
\Bigl( 2\chi T^0_0 +\delta T^0_0 \Bigr) - 6 \ch \Phi \vp \nonumber \\
&& - 3 \pp \vp - {\vec \nabla}^2 \chi + 3 \ch
\cp + 2 \Phi \phi^{\prime 2} - 2\cp \vp, \nonumber
\eea
\bea \label{eq112}
\partial_{i}\pp &+& \ch\partial_{i}\Phi \, = \\
&&  \frac{1}{2} e^{2\phi+2\lambda} \delta T_{0}^{i}+\partial_{i}\Phi
\vp-\partial_{i}\cp+\ch\partial_{i}\chi \nonumber 
\eea
\be \label{113}
\partial_{i}\partial_{j} \Bigl( \Phi-\Psi-2\chi \Bigr)=0 \;\;\;\;\;\; i \neq j,
\ee
\begin{widetext}
\bea
-2\ppp - 4\ch\pp  - 2\ch{^2}\Phi - 4\ch^{\prime}\Phi - 2\fp\ch \\
= \, e^{2 \phi+2\lambda} \Bigl( 2\chi T^{i}_{i}+\delta T^{i}_{i} \Bigr)
 - 4\Phi\vpp - 2\fp\vp  - 4\ch\Phi\vp -
4\pp\phi^{\prime} + 2\cpp  + 2\ch\cp + 
4\Phi\phi^{\prime 2} - 4\cp\vp, \nonumber
\eea
\end{widetext}
and
\bea \label{eq114}
-2 \Phi \phi^{\prime 2} &+& 2 \vp \cp+\Phi \vpp  
+ \frac{1}{2} \Phi^{\prime} \vp + 2 \ch \Phi \vp  \nonumber \\
&+& \frac{3}{2} \pp \vp  - \frac{1}{2}\cpp  
+\frac{1}{2}{\vec \nabla}^2 \chi - \ch \cp \nonumber \\
&=& \frac{1}{4}e^{2\phi + 2\lambda} \Bigl( 2\chi T + \delta T
\Bigr),
\eea
where $T\equiv T^{\mu}_{\mu}$ is the trace. The first equation
is the time-time equation, the second the space-time equation,
the next two the off-diagonal and diagonal space-space equations,
respectively, and the last one is the matter equation.

The times $t_i(k)$ when the metric perturbations were computed are
in the transition period between the Hagedorn phase and the radiation
phase of standard cosmology. Space is beginning to expand. In this
case, neither the terms containing the Hubble expansion rate nor
those containing the dilaton velocity vanish. 

We will first consider the case when the dilaton velocity is
negligible (we will come back to a discussion of when this is
a reasonable approximation), and then the case when the
dilaton velocity is important.

If the dilaton velocity is negligible, then the equations simplify
dramatically. We first note that at the time $t_i(k)$, the comoving
Hubble constant $\ch$ is of the same order of magnitude as $k$. 
Consider now, specifically, the time-time equation of motion
(\ref{eq111}). The terms containing $\ch$ and its derivative on
the left-hand side of this equation are of the same order of magnitude
as the first term. We will, therefore, neglect all terms containing
$\ch$. Hence, the equation simplifies to
\be \label{eq115}
{\vec \nabla}^2 \Psi 
\, = \, \frac{1}{2} e^{2\phi + 2\lambda}
\Bigl( 2\chi T^0_0 +\delta T^0_0 \Bigr) 
- {\vec \nabla}^2 \chi \, . 
\ee
Similarly, the perturbed dilaton equation simplifies to
\be \label{eq116}
- \frac{1}{2}\cpp + \frac{1}{2}{\vec \nabla}^2 \chi \, = \, 
\frac{1}{4}e^{2\phi + 2\lambda} \Bigl( 2\chi T + \delta T \Bigr),
\ee
From the latter equation, it follows that the Poisson spectrum
of $\delta T$ induces a Poisson spectrum of the dilaton fluctuation
$\chi$. Subtracting (\ref{eq116}) from (\ref{eq115}) and keeping
in mind that the background pressure vanishes in the Hagedorn
phase yields
\be
{\vec \nabla}^2 \Psi \, = \, \frac{1}{2} e^{2\phi + 2\lambda}
\Bigl( \delta T^0_0 - \delta T \Bigr) - \cpp
\ee
from which it follows that the induced spectrum of $\Psi$ will
be scale-invariant
\be
P_{\Psi} \, \sim \, k^0 \, .
\ee
It is easy to check that the other equations of motion are
consistent with this scaling. 

If the Hagedorn phase is modeled by the equations (\ref{E1} - \ref{E3}), 
then the dilaton velocity is not
negligible, since it is related to the energy density via the
constraint equation (\ref{E1}):
\be
{\dot \phi}^2 \, = \, {1 \over 4} e^{\varphi} E \, .
\ee
Thus, the terms containing the dilaton velocity are as important as the 
other terms in the Hagedorn phase. Now, the prescription of
\cite{ABV,BNPV2} was to use the constraint equation at the time
$t_i(k)$ when the scale $k$ exits the Hubble radius. This time is
towards the end of the Hagedorn phase. However, in the context of
our action, Eq. (\ref{E1}) always holds. The right-hand side
of this equation must be large since it gives the (square of the)
Hubble expansion rate at the beginning of the radiation phase. At
the time $t_i(k)$, then, for scales which are large compared to
the Hubble radius at the beginning of the radiation phase, the 
value of the terms containing ${\dot \lambda}$ in (\ref{E1}) are
negligible, and hence the dilaton velocity term is non-negligible. 

Let us now compute the induced metric fluctuations in
the Hagedorn phase taking into account the terms depending on the
dilaton velocity. The time-dependence of the dilaton introduces
a critical length scale into the problem, namely the inverse
time scale of the variation of the dilaton. Translated to the
Einstein frame, this length is the Einstein frame Hubble radius (this
radius is, up to a numerical constant, identical to the forward light
cone computed beginning at the time of the dilaton singularity).
On smaller scales, the dilaton-dependent terms in the time-time
Einstein constraint equation (\ref{eq111}) are negligible, 
Eq. (\ref{eq111})
reduces to the Poisson equation (\ref{Poisson}) and we conclude that
the Poisson spectrum of the stringy matter induces a flat
spectrum for the metric potential $\Psi$. 

On larger scales, however,
it is the dilaton-dependent terms in (\ref{eq111}) which dominate.
If we insist on the view that it is the string gas matter fluctuations
which seed all metric fluctuations, then we must take all terms
independent of the string sources to the left-hand side of the
equations of motion. If we do this and keep the terms on the left-hand
side of the equations which dominate in a gradient expansion, then
the time-time equation becomes
\be \label{eq117}
3 \pp \vp - 2 \Phi \phi^{\prime 2} + 2\cp \vp 
\, = \,  \frac{1}{2} e^{2\phi + 2\lambda}
\Bigl( 2\chi T^0_0 +\delta T^0_0 \Bigr) \, ,
\ee
and the analogous approximation scheme applied to the dilaton
equation yields
\bea \label{eq118}
-2 \Phi \phi^{\prime 2} &+& 2 \vp \cp +\Phi \vpp  
+ \frac{1}{2} \Phi^{\prime} \vp + \frac{3}{2} \pp \vp  - \frac{1}{2}\cpp 
\nonumber \\ 
&=& \frac{1}{4}e^{2\phi + 2\lambda} \Bigl( 2\chi T + \delta T \Bigr) \, .
\eea

Inspection of (\ref{eq118}) shows that a Poisson spectrum of
$\delta T$ will induce a Poisson spectrum of $\chi$. 
Subtracting two times (\ref{eq118}) from (\ref{eq117}) shows
that, given a Poisson spectrum of $\chi$, the resulting equation
is no longer consistent with a scale-invariant spectrum for $\Psi$
and $\Phi$, since terms which have a scale-invariant spectrum
would remain on the left-hand side of the equation. Hence, we
conclude that the induced spectrum of $\Phi$ and $\Psi$ will
also be Poisson:
\be
P_{\Psi} \, \sim \, k^4 \, .
\ee

The above conclusion is consistent with the Traschen Integral
constraints \cite{Traschen} which state that in the absence
of initial curvature fluctuations, motion of matter cannot produce
perturbations with a spectrum which is less red than Poisson on
scales larger than the horizon. This view is consistent with the
fact that on small scales, the spectrum is scale-invariant: on
small scales it is possible to move around matter by thermal fluctuations
to produce new curvature perturbations.

As we have stressed earlier, however, the equations (\ref{E1} - \ref{E3})
are definitely not applicable early in the Hagedorn phase, namely in
the strong coupling Hagedorn phase. In that phase, the dilaton is fixed, 
and thus
the arguments of \cite{ABV} imply the presence of scale-invariant
metric fluctuations seeded by the string gas perturbations 
\footnote{However, in this phase one must reconsider the 
computation of the
string thermodynamic fluctuations, since our analysis implicitly
assumes weak string coupling.}. These
fluctuations will persist in the phase in which the dilaton is
rolling (the fluctuations cannot suddenly decrease in magnitude).
Hence, we believe that the conclusions of \cite{ABV} are robust.

\section{Cosmological Perturbations in the Einstein Frame}

From the point of view of the Einstein frame, the scales
are super-Hubble during the phase of dilaton rolling. 
How is this consistent with
their sub-Hubble nature from the point of view of the
string frame? The answer is that, whereas the causal structure
of space-time (and thus concepts like horizons) are frame-independent,
the Hubble radius depends on the frame. The physical meaning of the
Hubble radius is that it separates scales on which matter oscillates
(sub-Hubble) from scales where the matter oscillations are frozen in
(super-Hubble). Matter which is coupled minimally to gravity in the
string frame feels the string frame Hubble radius
\footnote{Note, however, that the dilaton coupling to stringy
matter can produce friction effects which are similar to Hubble
friction.}, matter which
is minimally coupled to gravity in the Einstein frame feels
the Einstein frame Hubble radius. Strings couple minimally
to gravity in the string frame and hence feel the string frame
Hubble radius.

If we take into account the presence of the strong coupling Hagedorn phase,
then it becomes possible, also in the Einstein frame, to study
the generation of fluctuations. During the strong coupling Hagedorn phase,
the dilaton is fixed and hence the fluctuation equations are
those of Einstein gravity. Since scales of cosmological interest
today are sub-Hubble during this phase, a scale-invariant
spectrum of metric fluctuations is induced by the string gas
fluctuations, as discussed in the previous section. If the
strong coupling Hagedorn phase is long in duration, then a scale-invariant
spectrum can be induced    consistent with the Traschen integral
constraints. Note that the a long duration of the strong coupling
phase is required in order to justify the assumption of thermal
equilibrium on the scales we are interested in.

We can also obtain the Einstein frame initial conditions by
conformally transforming the initial conditions obtained
in the string frame. The transformation of the perturbation
variables in straightforward:
\bea
{\tilde \Psi} \, &=& \, \bigl( \Psi + \chi \bigr)  
\label{transf1} \\
{\tilde \Phi} \, &=& \, \bigl( \Phi - \chi \bigr) 
\label{transf2} \\
{\tilde \chi} \, &=& \, 2 \chi \, , \label{transf3} 
\eea
where, as before, tilde signs indicate quantities in the Einstein
frame.

Note, in passing, that the string frame off-diagonal spatial
equation of motion (\ref{113}) immediately implies that
\be
{\tilde \Psi} \, = \, {\tilde \Phi} \, ,
\ee
which is the well-known result for Einstein frame fluctuations
in the absence of matter with anisotropic stress.

Given the transformation properties (\ref{transf1} - \ref{transf3})
of the fluctuation variables, it is obvious that the conclusions
about the initial power spectra of the fluctuations variables are
the same as in the string frame: if the dilaton velocity can be
neglected, the spectrum of $\Phi$ and $\Psi$ is scale-invariant,
if the dilaton velocity is important, the spectra of these variables
are Poisson. 

Whereas setting the initial conditions for the fluctuations may look
more ad hoc in the Einstein frame, the evolution of the perturbations
is easier to analyze since we can use all of the intuition and results
developed in the context of fluctuations in general relativity.
In particular, we can use the Deruelle-Mukhanov \cite{DM} matching
conditions to determine the fluctuations in the post-Hagedorn radiation 
phase of standard cosmology from those at the end of the Hagedorn phase.
The Deruelle-Mukhanov conditions are generalization to space-like
hyper-surfaces of the Israel matching conditions \cite{Israel} which
state that the induced metric and the extrinsic curvature need to be
the same on both sides of the matching surface. Applied to the case
of cosmological perturbations, the result 
\cite{DM} is that (in terms of longitudinal
gauge variables) both ${\tilde \Phi}$ and $\zeta$ need to be continuous
\footnote{Note that the application of Israel matching conditions is, in
our case, well justified. The concerns raised in \cite{Durrer} regarding
the application of the matching conditions in the base of the
Pre-Big-Bang \cite{PBB} and Ekpyrotic/Cyclic \cite{KOST} scenarios do
not apply since in our case the matching conditions are satisfied at 
the level of the background solution.}
where $\zeta$ is defined as \cite{BST,BK,Lyth}
\be
\zeta \, \equiv \, {\tilde \Phi} + 
{{\ch} \over {\ch^2 - \ch^{\prime}}}
\bigl( {\tilde \Phi}^{\prime} + \ch {\tilde \Phi} \bigr) \, ,
\ee
where here $\ch$ the Einstein frame Hubble expansion rate with
respect to conformal time.

In the Einstein frame, the universe is radiation-dominated both
before and after the transition. Hence, in both phases the dominant mode
of the equation of motion for ${\tilde{\Phi}}$ is a constant. The constant
mode in the phase ${\tilde{t}} < {\tilde{t}}_R$ couples dominantly to the constant mode in the phase ${\tilde{t}} > {\tilde{t}}_R$. 
The initial value of the spectrum of
${\tilde{\Phi}}$ will seed both the constant and the decaying mode of 
${\tilde{\Phi}}$ with comparable strengths and with the same spectrum. 
Hence, the late-time value of ${\tilde{\Phi}}$ is
given, up to a factor of order unity, by the initial value
of ${\tilde{\Phi}}$ at the time ${\tilde{t}}_i(k)$
\be
P_{\tilde{\Phi}}(k, t) \, \simeq P_{\tilde{\Phi}}(k, {\tilde{t}}_i(k))\, 
\sim \, k^0 \,\,\,\, {\tilde{t}} >> {\tilde{t}}_R \, .
\ee

\section{Discussion and Conclusions}

In this paper we have recast string gas cosmology in the Einstein
frame rather than in the string frame in which the analysis usually
takes place. Our analysis sheds new light on several important
cosmological issues.
 
At the level of the background evolution, it becomes clear that
solutions of the dilaton gravity equations (\ref{E1} - \ref{E3})
with decreasing dilaton contain an initial singularity.
From the point of view of the Einstein frame,
there is an initial curvature singularity which follows from the
presence of a singularity in the dilaton field in the string frame.
Obviously, however, these solutions are not applicable at very early
times since they correspond to times when the string theory
is strongly coupled.   
Hence, there must be, prior to the phase of rolling dilaton, a
strong coupling Hagedorn phase in which
both the size of space and the dilaton must be quasi-static. Provided
that this phase lasts sufficiently long, thermal equilibrium over
all scales relevant to current observations can be established.

Note that during the period when the solutions of (\ref{E1} - \ref{E3})
have a rolling dilaton, then in the Einstein frame the 
expansion of space never accelerates. The evolution corresponds
to that of a radiation-dominated universe. The Einstein frame Hubble
radius increases linearly in time throughout. However, the presence
of the strong coupling Hagedorn phase can solve the horizon problem 
in the sense
of making the comoving horizon larger than the comoving scale
corresponding to our currently observed universe. 
  
How to model the strong coupling Hagedorn phase now becomes a crucial question for string gas cosmology \footnote{See also
\cite{Riotto} for an interesting discussion of these issues.}. 
Since the singularity of string gas cosmology is (from the point
of view of the string frame) associated with the dilaton becoming
large, and thus with string theory entering a strongly coupled
phase, it is interesting to conjecture that a process like tachyon
condensation \cite{tachyon} will occur       and  resolve the
singularity (like in the work of \cite{Eva}). If this phase
lasts for a long time, it will produce a large space in thermal equilibrium. 

There are other possible scenarios in which there is a precursor
phase of the rolling dilaton period which establishes thermal
equilibrium on large scales. One such possibility
was discussed in \cite{Natalia} and makes use of a pre-Hagedorn
phase in which the extra spatial dimensions initially expand, driven
by a gas of bulk branes. The resulting increase in the energy stored
in the branes leads to the increase in size and entropy which
solves both the horizon and entropy problems. Once the extra
spatial dimensions have contracted again to the string scale,
the size of our three spatial dimensions can be macroscopic
while the temperature of matter is of string scale.

Another possibility to obtain a solution to the horizon problem
and to justify thermal equilibrium over large scales is to
invoke a bouncing cosmology such as obtained in the context of
higher derivative gravity models in \cite{Biswas} (see also
\cite{MBS} for an earlier construction). The phase of
contraction could produce the high densities required to form
a string gas with the necessary requirements.

We have also studied the mechanism for the generation of fluctuations
proposed in \cite{ABV,BNPV2} in more detail, both from the point
of view of the string frame and the Einstein frame. We have
shown that a small value of the dilaton velocity in the Hagedorn phase
is required in order that 
the string thermodynamic fluctuations are able to generate a
scale-invariant spectrum of cosmological fluctuations. If dilaton velocity
terms are important, then a Poisson spectrum is produced.
If the dilaton velocity is negligible, and if thermal equilibrium on
the scales of interest can be justified, then a scale-invariant spectrum
of metric fluctuations is induced. 

\centerline{\bf Acknowledgements}

Two of us (R.B. and A.N.) are grateful to Lev Kofman, 
Andrei Linde and Slava Mukhanov for
detailed discussions, and for emphasizing the need to analyze the
dynamics of string gas cosmology completely in the Einstein frame.
R.B. thanks Misao Sasaki and the Yukawa Institute for Theoretical Physics for hospitality during the time when this project was initiated.
We wish to thank Misao Sasaki, Scott Watson and in particular Cumrun Vafa
for fruitful conversations. 
J.S. would like to thank the Department of Physics,
McGill University for kind hospitality during the course of this work.
The research of R.B. and S.P. is
supported by an NSERC Discovery Grant, by the Canada Research
Chairs program, and by an FQRNT Team Grant. The work of
D.E. is supported in part by PPARC. S.K. is supported by a 
JSPS Postdoctoral Fellowship for Research Abroad.
The research of J.K. at Perimeter Institute is supported in part by the 
Government of Canada through NSERC and by the Province of Ontario through 
MEDT. The work of A.N. is supported in part by
NSF grant PHY-0244821 and DMS-0244464.
The work of J.S. is supported by
the Japan-U.K. Research Cooperative Program, the Japan-France Research
Cooperative Program, and the Grant-in-Aid for  Scientific
Research Fund of the Ministry of Education, Science and Culture of Japan
No.18540262 and No.17340075. 



\end{document}